\documentclass[prd,aps,showpacs,nofootinbib,tightenlines]{revtex4}
\usepackage{amsmath}
\usepackage{amssymb}
\usepackage{epsfig}
\usepackage{graphicx}
\begin{document}

\newcommand{\beq}{\begin{eqnarray}}
\newcommand{\eeq}{\end{eqnarray}}

\newcommand{\non}{\nonumber\\ }
\newcommand{\rmt}{ {\rm T}}
\newcommand{\psl}{ p\!\!\!\slash}
\newcommand{\qsl}{ q \hspace{-2.0truemm}/ }

\newcommand{\epsl}{ \epsilon\!\!\!\slash}
\newcommand{\nsl}{ n \hspace{-2.2truemm}/ }
\newcommand{\vsl}{ v \hspace{-2.2truemm}/ }

%%%%%%%%%%%%%%%%%%%
\def \ctp{ {Commun. Theor. Phys. }}
\def \epjc{ {Eur. Phys. J. C }}
\def \jhep{ {J. High Energy Phys. }}
\def \jpg{ {J. Phys. G }}
\def \npb{ {Nucl. Phys. {\bf B}}}
\def \plb{ {Phys. Lett. B }}
\def \prd{ {Phys. Rev. D }}
\def \prl{ {Phys. Rev. Lett. }}
\def \ptp{ {Prog. Theor. Phys. }}
\def \zpc{ {Z. Phys. C }}
\def \cpc{ {Chin. Phys. C }}

%%%%%%%%%%%%%%%%%%%%%%%%%%%%%%%%%%%%%%%%%%%%

%%%%%%%%%%%%%%%%%%%%%%%%%%%%%%%%%%%%%%%%%%%%
\title{$S$-wave resonance contributions to the $B^0_{(s)}\to J/\psi\pi^+\pi^-$
and $B_s\to\pi^+\pi^-\mu^+\mu^-$ decays}
\author{Wen-Fei Wang$^1$}\email{wangwf@ihep.ac.cn}
\author{Hsiang-nan Li$^2$}\email{hnli@phys.sinica.edu.tw}
\author{Wei Wang$^{3,4}$}\email{wei.wang@sjtu.edu.cn}
\author{Cai-Dian L\"u$^{1,4}$}\email{lucd@ihep.ac.cn}
\affiliation{$^1$Center for Future High Energy Physics, Institute of High Energy Physics,\\
Chinese Academy of Sciences, Beijing 100049, China,}
\affiliation{$^2$Institute of Physics, Academia Sinica, Taipei, Taiwan 115, Republic of China,}
\affiliation{$^3$INPAC, Shanghai Key Laboratory for Particle Physics and Cosmology, Department of Physics and Astronomy, Shanghai Jiao Tong University, Shanghai, 200240,   China,}
\affiliation{$^4$State Key Laboratory of Theoretical Physics, Institute of Theoretical Physics, Chinese Academy of Sciences, Beijing 100190, China}

\date{\today}
%-----------------------------------------------------%
\begin{abstract}
We study $S$-wave resonance contributions to the $B^0_{(s)}\to J/\psi\pi^+\pi^-$ and
$B_s\to\pi^+\pi^-\mu^+\mu^-$ decays in the perturbative QCD framework by
introducing two-hadron distribution amplitudes for final states.
The Breit$-$Wigner formula for the $f_0(500)$, $f_0(1500)$, and $f_0(1790)$ resonances and
the Flatt\'e model for the $f_0(980)$ resonance are adopted to parametrize the timelike
scalar form factors in the two-pion distribution amplitudes, which include both
resonant and nonresonant contributions. The resultant branching fraction and differential
branching fraction in the pion-pair invariant mass for each resonance channel are
consistent with experimental data. The determined $S$-wave two-pion
distribution amplitudes, containing the information of both resonant and nonresonant
rescattering phases, can be employed to predict direct $CP$ asymmetries of other three-body
hadronic $B$ meson decays in various localized regions of two-pion phase space.

\end{abstract}

\pacs{13.20.He, 13.25.Hw, 13.30.Eg}

\maketitle
%------------------------------------------------------%
\section{INTRODUCTION}

After the LHCb Collaboration measured sizable direct $CP$ asymmetries in localized regions
of phase space~\cite{Aaij:2013sfa,Aaij:2013bla,Nasteva:2013bta,deMiranda:2013kg},
three-body hadronic $B$ meson decays have attracted a lot of attention recently.
To calculate direct $CP$ asymmetries in the kinematic region with two final-state hadrons
almost collimating to each other, we have proposed a theoretical framework
based on the perturbative QCD (PQCD) approach with the crucial nonperturbative
input of two-hadron distribution amplitudes~\cite{Chen:2002th,Chen:2004az}.
The typical PQCD factorization formula for a $B\to h_1h_2h_3$ decay
amplitude is written as~\cite{Chen:2002th,Chen:2004az}
\beq
\mathcal{A}=\phi_B\otimes H\otimes \phi_{h_1h_2}\otimes\phi_{h_3},
\eeq
where the leading-order hard kernel $H$ contains only one hard gluon as in the
formalism for two-body $B$ meson decays, and
the $B$ meson ($h_1$-$h_2$ pair, $h_3$ meson) distribution amplitude $\phi_B$
($\phi_{h_1h_2}$, $\phi_{h_3}$) absorbs nonperturbative dynamics in the process.
The analysis of three-body hadronic $B$ meson decays is then
reduced to that of two-body ones. Fitting the timelike form factors and
rescattering phases contained in the two-pion distribution amplitudes to relevant
experimental data, we have been able to make predictions for the direct $CP$
asymmetries in the $B^\pm \to \pi^+ \pi^- \pi^\pm$ and
$\pi^+\pi^-K^\pm$ modes, which are consistent with the LHCb data in
the localized region of the pion-pair invariant mass squared
$m^2_{\pi^+\pi^-\rm low} < 0.4$ GeV$^2$~\cite{Wang:2014ira}.

A limitation of the study in Ref.~\cite{Wang:2014ira} is that only nonresonant
contributions to the timelike form factors were included, so it cannot
be extended to regions involving intermediate resonances. In this paper we will
propose parameterizations of the complex timelike form factors, which include
both resonant and nonresonant contributions, taking the $S$-wave two-pion distribution
amplitudes as an example. The LHCb data of $B_s$ meson decays through $S$-wave
resonances, such as $B_s\to J/\psi f_0(980)[f_0(980)\to\pi^+\pi^-]$ and
$B_s\to f_0(980)[f_0(980)\to\pi^+\pi^-]\mu^+\mu^-$, are then adopted to
fix involved parameters. It will be demonstrated that the resultant branching fraction
and differential branching fraction in the pion-pair invariant mass for each $S$-wave
resonance channel agree with the data. It implies
that the PQCD approach~\cite{Keum:2000ph,Keum:2000wi,Lu:2000em} is an appropriate
framework for analyzing three-body hadronic $B$ meson decays. The determined $S$-wave
two-pion distribution amplitudes, containing the information of both resonant and
nonresonant rescattering phases, can be employed to predict direct $CP$ asymmetries
of other three-body hadronic $B$ meson decays like $B^\pm\to K^\pm\pi^+\pi^-$ and
$B^\pm\to \pi^\pm\pi^+\pi^-$ in various localized regions of two-pion phase space.
The same formalism will be applied to the extraction of the $P$-wave two-pion
distribution amplitudes from, for instance, the Belle data~\cite{Garmash:2005rv} in the future.

The $B_s\to J/\psi \pi^+\pi^-$ decay was first measured by the LHCb Collaboration~\cite{Stone:2008ak},
one of whose channels $B_s\to J/\psi f_0(980)[f_0(980)\to\pi^+\pi^-]$
with a $CP$-odd final eigenstate can be used to extract the mixing-induced $CP$
violation phase $\phi_s$. A modified Dalitz-plot analysis~\cite{LHCb:2012ae} indicated
that the $f_0(980)$ and $f_0(1370)$ resonances give the dominant intermediate
contributions, about $70\%$ and $21\%$ of the total decay fraction, respectively.
It was shown based on $3$ fb$^{-1}$ of integrated luminosity~\cite{Aaij:2014emv} that
the $S$-wave resonances $f_0(980)$, $f_0(1500)$ and a so-called $f_0(1790)$ state, along
with tiny components from the $D$-wave resonances $f_2(1270)$ and $f^\prime_2(1525)$,
well describe the $B_s\to J/\psi \pi^+\pi^-$ data. The Cabibbo- and color-suppressed
decay $B^0\to J/\psi \pi^+\pi^-$ was first observed by the $BABAR$ Collaboration with the
branching ratio $(4.6\pm0.7\pm0.6)\times10^{-5}$, to which the $B^0\to J/\psi \rho^0$
channel was found to contribute $(1.6\pm0.6\pm0.4)\times10^{-5}$~\cite{Aubert:2002vb},
and $(2.7\pm0.3\pm0.17)\times10^{-5}$ in a later improved measurement~\cite{Aubert:2007xw}.
The LHCb Collaboration confirmed these results by investigating resonant components in the
$B^0\to J/\psi \pi^+\pi^-$ mode thoroughly~\cite{Aaij:2014siy,Aaij:2013zpt}: the
intermediate $\rho^0$ meson dominates, contributing about $63\%$ of the total decay
fraction~\cite{Aaij:2014siy}, and the $S$-wave resonance $f_0(500)$ is
the next, giving about $22\%$~\cite{Aaij:2014siy}. In contrast to the sizable $f_0(500)$
component, another $S$-wave state $f_0(980)$ was not seen in the
$B^0\to J/\psi \pi^+\pi^-$ decay~\cite{Aaij:2014siy,Aaij:2013zpt}.

The first observation of the $B_s\to \pi^+\pi^-\mu^+\mu^-$ decay and the first
evidence of the $B^0\to \pi^+\pi^-\mu^+\mu^-$ decay have been announced by the LHCb
Collaboration~\cite{Aaij:2014lba}, with the branching ratios
$(8.6\pm1.5\pm 0.7\pm0.7)\times10^{-8}$ and $(2.11\pm0.51\pm 0.15\pm0.16)\times10^{-8}$,
respectively. Within the measured range of the pion-pair invariant mass 0.5-1.3 GeV,
the $B_s\to f_0(980)\mu^+\mu^-$ ($B^0\to \rho(770)^0\mu^+\mu^-$) channel
is expected to dominate in the former (latter), contributing
$(8.3\pm1.7)\times10^{-8}$ ($(1.98\pm0.53)\times10^{-8}$). Hence, the
$B_s\to \pi^+\pi^-\mu^+\mu^-$ data also provide a useful input for the determination of
the $S$-wave two-pion distribution amplitudes. It will be shown that
they are crucial for fixing the involved Gegenbauer coefficient, to which
the $B_s\to J/\psi \pi^+\pi^-$ data are less sensitive. For an independent study
of the $B^0\to \pi^+\pi^-\mu^+\mu^-$ decay in the
light-cone sum rules, refer to Ref.~\cite{Wang:2015paa}. The
$B^0\to \rho(770)^0[\rho(770)^0\to\pi^+\pi^-]\mu^+\mu^-$ decay is applicable to
the extraction of the $P$-wave two-pion distribution amplitudes stated before.

The factorization formalism of the $B^0_{(s)}\to J/\psi\pi^+\pi^-$
and $B_s\to\pi^+\pi^-\mu^+\mu^-$ decays with the inputs of the $S$-wave two-pion
distribution amplitudes is established in Sec.~II. The associated
timelike scalar form factors are parameterized in terms of the Breit-Wigner (BW)
model for the $f_0(500)$, $f_0(1500)$ and $f_0(1790)$ resonances and
the Flatt\'e model for the $f_0(980)$ resonance. In Sec.~III
the relative strengths and strong phases among the above resonances are
determined by fitting the factorization formulas to the LHCb data. It will be
highlighted that the obtained differential branching fraction in the pion-pair
invariant mass for each resonance channel matches well the experimental
one. Especially, our results for the $B_s\to J/\psi\pi^+\pi^-$ decay favor
the "Solution~I" set of branching fractions for various resonances presented
by the LHCb Collaboration~\cite{Aaij:2014emv}. Section~IV contains the conclusion.
The explicit PQCD factorization formulas for all the
decay amplitudes are collected in the Appendix.

\section{DECAY AMPLITUDES}

\subsection{$B^0_{(s)}\to J/\psi\pi^+\pi^-$ Decay}

\begin{figure}[tbp]
\vspace{-1cm}
\centerline{\epsfxsize=16cm \epsffile{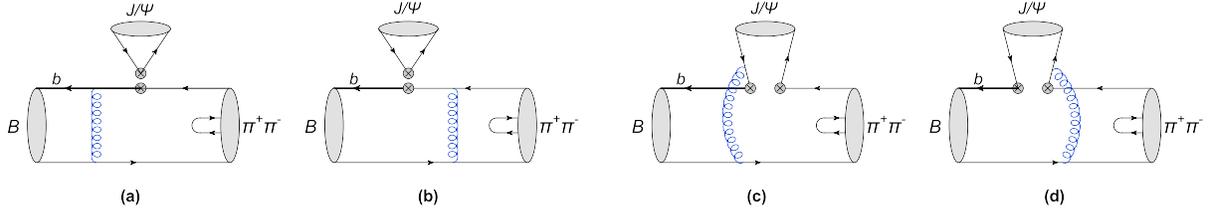}}
\caption{Typical Feynman diagrams for the three-body decays $B^0_{(s)}\to J/\psi\pi^+\pi^-$,
where $B$ stands for the $B^0$ or $B_s$ meson, and $\otimes$ denotes the weak vertex.}
\label{fig-fig1}
\end{figure}

The $B^0_{(s)}\to J/\psi\pi^+\pi^-$ modes mainly proceed via quasi-two-body channels
containing scalar or vector resonant states as argued
in Refs.~\cite{Chen:2002th,Wang:2014ira}.
For discussions on the $B_s\to J/\psi f_0(980) [f_0(980)\to \pi^+\pi^-]$ channel,
refer to Refs.~\cite{Colangelo:2010bg,Leitner:2010fq,Colangelo:2010wg,Li:2012sw,Fleischer:2011au}.
The chiral unitary approach including final state interaction~\cite{Liang:2014tia,Bayar:2014qha,Xie:2014gla}
was employed to investigate the $B^0_{(s)}\to J/\psi\pi^+\pi^-$ decays recently.
The $B^0\to J/\psi \pi^+\pi^-$ branching ratio was calculated in the QCD-improved
factorization approach, where a two-meson distribution
amplitude for the pion pair and final state interaction were considered~\cite{Sayahi:2013tza}.
In this section we derive the PQCD factorization formulas for the $B^0_{(s)}\to J/\psi\pi^+\pi^-$
decays with the inputs of the $S$-wave two-pion distribution amplitudes. It has been
postulated that the leading-order hard kernel for three-body $B$ meson decays contains only
one hard gluon exchange as depicted in Fig.~\ref{fig-fig1}, where the $B^0$ or $B_s$ meson
transits into a pair of the $\pi^+$ and $\pi^-$ mesons through an intermediate resonance.
Figures~\ref{fig-fig1}(a) and \ref{fig-fig1}(b) represent the factorizable contribution, and
Figs.~\ref{fig-fig1}(c) and \ref{fig-fig1}(d) represent the spectator contribution.

%%%%%%%%%%%%%%%%%%%%%%%%%%%%%%%%%%%%%%%%%%%%%%%%%%
%%%%%%%%%%%%%%%%%%%%%%%%%%%%%%%%%%%%%%%%%%%%%%%%%%

In the light-cone coordinates, the $B^0_{(s)}$ meson
momentum $p_{B}$, the total momentum of the two pions, $p=p_1+p_2$, and the $J/\psi$ momentum
$p_3$ are chosen as
\beq\label{mom-pBpp3}
p_{B}=\frac{m_{B}}{\sqrt2}(1,1,0_\rmt),~\quad p=\frac{m_{B}}{\sqrt2}(1-r^2,\eta,0_\rmt),~\quad
p_3=\frac{m_{B}}{\sqrt2}(r^2,1-\eta,0_\rmt),
\eeq
where $m_{B}$ denotes the $B^0_{(s)}$ meson mass, $m_{J/\psi}$ is the $J/\psi$ meson mass,
and the variable $\eta$ is defined as
\beq\label{def-eta}
\eta=\frac{\omega^2}{(1-r^2)m^2_{B}},
\eeq
with the mass ratio $r=m_{J/\psi}/m_{B}$ and the invariant mass squared $\omega^2=p^2$ of the
pion pair. Define $\zeta=p^+_1/p^+$ as the $\pi^+$ meson momentum fraction, in terms of which
the other kinematic variables of the two pions are expressed as
\beq
p^-_1=(1-\zeta)\eta\frac{m_{B}}{\sqrt2},
\quad p^+_2=(1-\zeta)(1-r^2)\frac{m_{B}}{\sqrt2}, \quad p^-_2=\zeta\eta\frac{m_{B}}{\sqrt2}.
\eeq
The momenta for the spectators in the ${B^0_{(s)}}$ meson, the pion pair, and the $J/\psi$ meson read as
\beq\label{mom-B-k}
k_{B}&=&\left(0,\frac{m_{B}}{\sqrt2}x_{B},k_{BT}\right),\quad
k=\left(\frac{m_{B}}{\sqrt2}z(1-r^2),0,k_\rmt\right),\non
k_3&=&\left(\frac{m_{B}}{\sqrt2}r^2x_3,\frac{m_{B}}{\sqrt2}(1-\eta)x_3,k_{3\rmt}\right),
\eeq
where the momentum fractions $x_{B}$, $z$, and $x_3$ run between zero and unity.

We introduce the distribution amplitudes for the pion pair~\cite{Mueller:1998fv,Diehl:1998dk,Polyakov:1998ze},
\begin{eqnarray}
\Phi_{v\nu}^I(z,\zeta,\omega^2)&=&\frac{1}{2\sqrt{2N_c}}\int \frac{dy^-}{2\pi}
e^{-izp^+y^-}\langle\pi^+(p_1)\pi^-(p_2)|\bar\psi(y^-)\gamma_\nu T
\psi(0)|0\rangle,
\label{pa}\\
\Phi_s^I(z,\zeta,\omega^2)&=&\frac{1}{2\sqrt{2N_c}}\frac{p^+}{w}
\int \frac{dy^-}{2\pi}
e^{-izp^+y^-}\langle\pi^+(p_1)\pi^-(p_2)|\bar\psi(y^-)T
\psi(0)|0\rangle,
\label{ps}\\
\Phi_{t\nu}^I(z,\zeta,\omega^2)&=&\frac{1}{2\sqrt{2N_c}}\frac{f_{2\pi}^\perp}{w^2}
\int \frac{dy^-}{2\pi}
e^{-izp^+y^-}\langle\pi^+(p_1)\pi^-(p_2)|\bar\psi(y^-)
i\sigma_{\mu\nu}n_-^\mu  T\psi(0)|0\rangle,
\label{pt}
\end{eqnarray}
where $T=1/2$ ($T=\tau^3/2$) is for the isoscalar $I=0$ (isovector $I=1$)
state, $N_c$ the number of colors, $\psi$ the $u$-$d$ doublet, $z$ the momentum fraction
carried by the spectator $u$ quark, and $n_-=(0,1,{\bf 0}_T)$ a dimensionless vector.
For $I=1$, the $P$-wave is the leading partial wave, and
$\Phi_{v\nu=-}^{I=1}$ and $\Phi_{t\nu=\perp}^{I=1}$
contribute at twist-2. The other components $\Phi_{v\nu=\perp}^{I=1}$,
$\Phi_s^{I=1}$, and $\Phi_{t\nu=+}^{I=1}$ then contribute at
twist-3. The $\nu=+$, $-$, and $\perp$ components can be extracted by
contracting the above definitions with $p^+$, $n_-$, and $p_{iT}$,
respectively. For $I=0$, the S-wave is the leading partial wave, and
$\Phi_{v\nu=-}^{I=0}$ contributes at twist-2. The others $\Phi_{s}^{I=0}$
and $\Phi_{t\nu=+}^{I=0}$ contribute at twist-3. In the case with the pion
pair coming from the $s$ and $\bar s$ quarks, $\psi$ is replaced by
the $s$ quark field. A two-pion distribution amplitude can be related to the pion
distribution amplitude through the evaluation of the process
$\gamma\gamma^*\to \pi^+\pi^-$ at large invariant mass $w^2$~\cite{Diehl:1999ek,Diehl:2000uv}.
The extraction of the two-pion distribution amplitudes from the
$B\to \pi\pi l\bar\nu$ decay has been elaborated in Refs.~\cite{Maul:2001zn,Faller:2013dwa}.

The $S$-wave two-pion distribution amplitudes are collected into~\cite{Meissner:2013hya}
\begin{eqnarray}
\Phi_{\pi\pi}^{S-wave}=\frac{1}{\sqrt{2N_c}}\left[\psl\Phi_{v\nu=-}^{I=0}(z,\zeta,w^2)
+\omega\Phi_{s}^{I=0}(z,\zeta,w^2)+\omega(\nsl_+\nsl_--1)\Phi_{t\nu=+}^{I=0}(z,\zeta,w^2) \right],
\end{eqnarray}
$n_+=(1,0,{\bf 0}_T)$ being a dimensionless vector. Their asymptotic forms are parameterized
as~\cite{Mueller:1998fv,Diehl:1998dk,Polyakov:1998ze}
\begin{eqnarray}
\Phi_{v\nu=-}^{I=0}=\frac{9F_{s}(w^2)}{\sqrt{2N_c}}a_2^{I=0}z(1-z)(1-2z),\quad
\Phi_{s}^{I=0}=\frac{F_{s}(w^2)}{2\sqrt{2N_c}},\quad
\Phi_{t\nu=+}^{I=0}=\frac{F_{s}(w^2)}{2\sqrt{2N_c}}(1-2z),
\end{eqnarray}
with the timelike scalar form factor $F_{s}(w^2)$ and the Gegenbauer coefficient $a_2^{I=0}$.
Following the analysis of the LHCb Collaboration~\cite{LHCb:2012ae,Aaij:2014emv,Aaij:2013zpt,Aaij:2014siy},
and motivated by the studies in Refs.~\cite{Hagler:2002nh,Hagler:2002nf,Pire:2008xe},
we introduce the $S$-wave resonances into the parametrization of $F_{s}(w^2)$,
so that both resonant and nonresonant contributions are included into the $S$-wave
two-pion distribution amplitudes. For the $s\bar s$ component in the $B_s\to J/\psi\pi^+\pi^-$
decay, we take into account $f_0(980), f_0(1500)$, and $f_0(1790)$ as in
Ref.~\cite{Aaij:2014emv}.  It was noticed by
the LHCb Collaboration that the above resonant states
lead to a fit better than in Ref.~\cite{LHCb:2012ae}, where
$f_0(980), f_0(1370)$, and $f_0(1500)$ were considered.
Note that there is a resonant state $f_0(1710)$, instead of $f_0(1790)$,
in Refs.~\cite{Beringer:1900zz,Agashe:2014kda} with the mass around $1.72$ GeV.

For the $f_0(1500)$ and $f_0(1790)$ contributions, the BW formula is employed
\beq
\frac{1}{m_S^2-\omega^2-im_S\Gamma_S(\omega^2)},
\eeq
where $m_S$ is the pole mass of the resonance, and the energy-dependent
width $\Gamma_S(\omega^2)$ for a $S$-wave resonance decaying into two pions is parameterized as
\beq
\Gamma_S(\omega^2)=\Gamma_S \frac{m}{\omega}
\left(\frac{\omega^2-4m_{\pi}^2}{m_S^2-4m_{\pi}^2} \right)^{\frac{1}{2}} F_R^2,
\eeq
with the pion mass $m_{\pi}$, the constant width $\Gamma_S$, and the Blatt-Weisskopf
barrier factor $F_R=1$ in this case~\cite{LHCb:2012ae,blatt-weisskopf}. The values of
$m_{f_0(1500)}=1.50$ GeV and $\Gamma_{f_0(1500)}=0.12$ GeV are taken for the mass and the
width of the $f_0(1500)$ meson, and
$m_{f_0(1790)}=1.81$ GeV and $\Gamma_{f_0(1790)}=0.32$ GeV
are taken for $f_0(1790)$~\cite{Aaij:2014emv,Beringer:1900zz,Agashe:2014kda}.
The BW formula does not work well for $f_0(980)$,
since this meson is close to the $K\bar K$ threshold. We then employ
the Flatt\'e model~\cite{Flatte:1976xv},
\beq
\frac{1}{m_{f_0(980)}^2-\omega^2-im_{f_0(980)}(g_{\pi\pi}\rho_{\pi\pi}+g_{KK}\rho_{KK})},
\eeq
with the pole mass $m_{f_{0}(980)}=0.97$
GeV~\cite{Ablikim:2004wn,Ambrosino:2005wk,Beringer:1900zz,Agashe:2014kda}. The coupling
constants $g_{\pi\pi}=0.167$ GeV and $g_{KK}=3.47g_{\pi\pi}$~\cite{Aaij:2014siy,Aaij:2014emv}
describe the $f_0(980)$ decay into the final states $\pi^+\pi^-$ and $K^+K^-$, respectively.
The phase space factors $\rho_{\pi\pi}$ and $\rho_{KK}$ read
as~\cite{Aaij:2013zpt,Flatte:1976xv,Aaij:2014emv}
\beq
\rho_{\pi\pi}=\frac23\sqrt{1-\frac{4m^2_{\pi^\pm}}{\omega^2}}
 +\frac13\sqrt{1-\frac{4m^2_{\pi^0}}{\omega^2}},\quad
\rho_{KK}=\frac12\sqrt{1-\frac{4m^2_{K^\pm}}{\omega^2}}
 +\frac12\sqrt{1-\frac{4m^2_{K^0}}{\omega^2}}.
\eeq

Considering the relative strengths and strong phases among different resonances,
we write the timelike scalar form factor associated with the $s\bar s$ component as
\beq
F_s^{s\bar s}(\omega^2)&=&
\frac{c_1m_{f_0(980)}^2e^{i\theta_1}}{m_{f_0(980)}^2-\omega^2-im_{f_0(980)}
  (g_{\pi\pi}\rho_{\pi\pi}+g_{KK}\rho_{KK})} \nonumber\\
  &&+\frac{c_2m_{f_0(1500)}^2e^{i\theta_2}}{m_{f_0(1500)}^2-\omega^2-im_{f_0(1500)}\Gamma_{f_0(1500)}(\omega^2)}\non
& &+\frac{c_3m_{f_0(1790)}^2e^{i\theta_3}}{m_{f_0(1790)}^2-\omega^2-im_{f_0(1790)}\Gamma_{f_0(1790)}(\omega^2)},
\label{ff-expr-ss}
\eeq
$c_i$ and $\theta_i$, $i=1$, 2, and 3, being tunable parameters.
The above parametrization decreases like $\omega^{-2}$ in the asymptotic limit, indicating
that it also includes the nonresonant contribution~\cite{Chen:2002th}.
For the $d\bar d$ component in the $S$-wave two-pion distribution amplitude, only the resonance
$f_0(500)$ or the so-called $\sigma$ meson in literature is needed.
It was observed that the $f_0(980)$ contribution to the $B^0\to J/\psi\pi^+\pi^-$ branching ratio
is less than 1\% compared with the $f_0(500)$ one~\cite{Aaij:2014siy}. That is,
the LHCb data~\cite{Aaij:2014siy,Aaij:2013zpt} disfavor the existence
of $f_0(980)$ in the $B^0\to J/\psi\pi^+\pi^-$ decay, rendering doubtful the
hypothesis of a four-quark state (with the $d\bar d$ content) for this meson.
The $f_0(500)$ contribution to the scalar form factor for the $d\bar d$ component
is parameterized as a BW model with the pole mass $m_{f_0(500)}=0.50$ GeV and the
width $\Gamma_{f_0(500)}=0.40$ GeV according to
Refs.~\cite{Aaij:2013zpt,Aaij:2014siy,Beringer:1900zz,Agashe:2014kda,Muramatsu:2002jp},
\beq
F_s^{d\bar d}(\omega^2)=\frac{c m_{f_0(500)}^2}{m_{f_0(500)}^2-\omega^2
-im_{f_0(500)}\Gamma_{f_0(500)}(\omega^2)},
\eeq
$c$ being a tunable parameter.

The differential branching ratio for the $B^0_{(s)}\to J/\psi\pi^+\pi^-$ decay is expressed
as~\cite{Beringer:1900zz,Agashe:2014kda}
\beq
\frac{d{\cal B}}{d\omega}=\tau_{B}\frac{\omega|\overrightarrow{p_1}|
|\overrightarrow{p_3}|}{4(2\pi)^3m^3_{B}}|{\cal A}|^2,
\label{expr-br}
\eeq
with the $B^0_{(s)}$ meson mean lifetime $\tau_{B}$.
The kinematic variables $|\overrightarrow{p_1}|$
and $|\overrightarrow{p_3}|$ denote the magnitudes of the $\pi^+$ and $J/\psi$
momenta in the center-of-mass frame of the pion pair,
\beq
 |\overrightarrow{p_1}|=\frac12\sqrt{\omega^2-4m^2_{\pi^\pm}}, \quad~~
 |\overrightarrow{p_3}|=\frac{1}{2\omega}
  \sqrt{\left[m^2_{B}-(\omega+m_{J/\psi})^2 \right]\left[m^2_{B}-(\omega-m_{J/\psi})^2 \right]}.
\eeq
The decay amplitudes ${\cal A}$ from Fig.~\ref{fig-fig1} with the
$S$-wave pion pair in the final state are collected in the Appendix.

%%%%%%%%%%%%%%%%%%%%%%%%%%%%%%%%%%%%%%%%%%%%%%%
\subsection{$B_s\to \pi^+\pi^- \mu^+\mu^-$ Decay}
%\label{sec:differentialdecaydistribution}
%%%%%%%%%%%%%%%%%%%%%%%%%%%%%%%%%%%%%%%%%%%%%%%

Next we establish the PQCD factorization formalism for the differential branching ratio
of the $B_s\to \pi^+\pi^-\mu^+\mu^-$ decay, which is generically a
four-body mode. This decay is governed by the effective Hamiltonian
 \begin{eqnarray}
 {\cal
 H}_{\rm{eff}}=
 -\frac{G_F}{\sqrt{2}}V_{tb}V^*_{ts}\sum_{i=1}^{10}C_i(\mu)O_i(\mu),
 \label{eq:Hamiltonian}
 \end{eqnarray}
where $O_i$ and $C_i$ are a four-quark or magnetic-penguin operator
and its corresponding Wilson coefficient, respectively, the renormalization
scale $\mu$ will be set to the $b$ quark mass $m_b$ below, $G_F$ is the Fermi constant,
and $V_{tb}$ and $V_{ts}$ are the CKM matrix elements.
Equation~(\ref{eq:Hamiltonian}) yields the effective operators for the
$b\to s\ell^+\ell^-$ transition
\begin{eqnarray}
 i {\cal M}(b\to
 s\ell^+\ell^-)&=& iN_1
 \bigg\{ ({C_9+C_{10}})[\bar sb]_{V-A}[\bar \ell\ell]_{V+A}
 +({C_9-C_{10}})[\bar sb]_{V-A}[\bar \ell\ell]_{V-A}  \nonumber\\
 && + 4C_{7L}m_b\frac{q^\mu}{q^2}[\bar s i\sigma_{\mu\nu}
 (1+\gamma_5)b][\bar \ell\gamma^\nu \ell]
 + 4C_{7R}m_b\frac{q^\mu}{q^2}[\bar s i\sigma_{\mu\nu}
 (1-\gamma_5)b][\bar \ell \gamma^\nu
 \ell]\bigg\},\label{eq:decay-amplitude-bsll-LR}
\end{eqnarray}
with $C_{7L}=C_7$, $C_{7R}=C_{7L}{m_s}/{m_b}$, $m_s$ being the $s$ quark
mass, the lepton-pair momentum $q$, and the coefficient
\begin{eqnarray}
N_1= \frac{ G_F} {4\sqrt 2} \frac{\alpha_{\rm em}}{\pi} V_{tb}V_{ts}^*.
\end{eqnarray}

The $B_s\to \pi^+\pi^-\mu^+\mu^-$ decay amplitude
is obtained by sandwiching Eq.~\eqref{eq:decay-amplitude-bsll-LR} between
the initial and final hadronic states, to which only the factorizable diagrams,
Figs.~\ref{fig-fig1}(a) and \ref{fig-fig1}(b), contribute.
The spinor products of $\bar s$ and $b$ are
then replaced by the hadronic matrix elements~\cite{Doring:2013wka,Meissner:2013pba}
\begin{eqnarray}
 \langle (\pi^+\pi^-)_S(p)|\bar s \gamma_\mu\gamma_5 b|\overline B_s (p_{B_s})
 \rangle  &=& -i  \frac{1}{\omega} \bigg\{ \bigg[P_{\mu}
 -\frac{m_{B_s}^2-\omega^2}{q^2} q_\mu \bigg] {\cal F}_{1}(q^2)
 +\frac{m_{B_s}^2-\omega^2}{q^2} q_\mu  {\cal F}_{0}(q^2)  \bigg\},
 \nonumber\\
 \langle (\pi^+\pi^-)_S(p)|\bar s \sigma_{\mu\nu} q^\nu \gamma_5 b|
 \overline B_s (p_{B_s})\rangle  &=& \bigg[ ({m_{B_s}^2-\omega^2}) q_\mu - q^2
 P_{\mu}\bigg]\frac{{\cal F}_T(q^2)}{\omega(m_{B_s}+\omega)},
 \label{eq:generalized_form_factors}
\end{eqnarray}
that define the $\overline B_s\to \pi^+\pi^-$
form factors ${\cal F}_{0}$, ${\cal F}_{1}$, and ${\cal F}_{T}$
with the sum of the momenta $P=p_{B_s}+p$. The explicit factorization formulas for
the above form factors are referred to the Appendix. Since the PQCD approach is
applicable only to the large recoil region, the extrapolation of the form factor
behavior to small recoil is necessary for the evaluation of the branching
ratio. Here, we adopt the dipole form for the
the $q^2$ dependence of the form factor ${\cal F}_i$, $i=0$, 1, and $T$,
\begin{eqnarray}
  {\cal F}_i(q^2)= \frac{ {\cal F}_i(0)}{1-a_iq^2/m_{B_s}^2+ b_i (q^2/m_{B_s}^2)^2},
 \end{eqnarray}
where the shape parameters $a_i$ and $b_i$, depending on $\omega^2$, will be obtained from the knowledge
in the range $q^2<10{\rm GeV}^2$.

A general differential decay width for the
$B\to h_1h_2\ell^+\ell^-$ mode with various partial wave contributions has been derived
using the helicity amplitudes in Refs.~\cite{Lu:2011jm,Li:2010ra}. In the
case with the dominant $S$-wave contribution, the angular distribution is written as
\begin{eqnarray}
 \frac{d^3\Gamma}{d\omega^2dq^2  d\cos\theta_\ell  }
 &=& \frac{3}{8}\Big[J_1^c   + J_2^c
 \cos(2\theta_\ell)  \Big],
\end{eqnarray}
with $\theta_\ell$ being the polar angle between the $\ell^-$ and $B_s$ moving directions in the
lepton-pair rest frame. The angular coefficients are given by
\begin{eqnarray}
J_1^c&=&   \bigg[   |{\cal A}^0_{L0}|^2+|{\cal A}^0_{R0}|^2
 +8  \hat m_\ell^2  | {\cal A}^0_{L0}{\cal A}^{0*}_{R0} | \cos(\delta_{L0}^0 -\delta_{R0}^0)
 +4 \hat m_\ell^2  |{\cal A}_t^0|^2 \bigg], \nonumber \\
 J_2^c  &=& -\beta_\ell^2   \bigg[   |{\cal A}^0_{L0}|^2+|{\cal A}^0_{R0}|^2    \bigg] ,
  \label{eq:simplified_angularCoefficients_S-wave}
\end{eqnarray}
with $\beta_\ell=\sqrt{1-4m_\ell^2/q^2}$ and $\hat m_\ell= m_\ell/\sqrt{q^2}$,
in which the subscript $t$ denotes the timelike component of a virtual state decaying
into a lepton pair, and the strong phases $\delta_{L0}^0$ and $\delta_{R0}^0$ associated
with the helicity amplitudes ${\cal A}_{L0}$ and
${\cal A}_{R0}$, respectively, vanish at leading order in the strong coupling $\alpha_s$.
The helicity amplitudes are expressed, in terms of the form factors, as
\begin{eqnarray}
 {\cal A}_{L/R,0}^0&=&\sqrt{N_2} \frac{i}{\omega}\Bigg[ (C_9\mp C_{10})
 \frac{\sqrt {\lambda}}{\sqrt{ q^2}} {\cal F}_1(q^2) +2(C_{7L}-C_{7R})
 \frac{\sqrt {\lambda }m_{B_s}}{\sqrt {q^2}(m_{B_s}+\omega)}{\cal F}_T(q^2) \Bigg],\nonumber\\
   {\cal A}_{t}^0
&=&  2 \sqrt{N_2}C_{10}   \frac{i}{\omega}\Bigg[  \frac{m_{B_s}^2-\omega^2}{\sqrt {q^2}} {\cal F}_0(q^2) \Bigg],
\end{eqnarray}
with the factor
\begin{eqnarray}
N_2= \frac{1}{16\pi^2}N_1 N_{\pi\pi} \sqrt{1-4m_{\pi}^2/\omega^2},\;\;\;\;
 N_{\pi\pi} = \sqrt{\frac{8}{3}} \frac{\sqrt {\lambda}
{q^2}\beta_\ell}{256\pi^3 m_{B_s}^3},
\end{eqnarray}
and the function $\lambda\equiv\lambda(m^2_{B_s},\omega^2,q^2)=(m^2_{B_s}-\omega^2-q^2)^2-4\omega^2q^2$.

%%%%%%%%%%%%%%%%%%%%%%%%%%%%%%%%%%%%%%%%%%%%%%%%%%%%
\section{Numerical Results}
\label{sec:numerics}
%%%%%%%%%%%%%%%%%%%%%%%%%%%%%%%%%%%%%%%%%%%%%%%%%%%%

In addition to the quantities that have been
specified before, we take the following inputs (in units of
GeV)~\cite{Wang:2014ira,Beringer:1900zz,Agashe:2014kda}
\beq
\Lambda^{(f=4)}_{ \overline{MS} }&=&0.250, \quad m_{B^0}=5.280, \quad m_{B_s}=5.367,
\quad m_{J/\psi}=3.097, \nonumber\\
m_{\pi^\pm}&=&0.140, \quad m_{\pi^0}=0.135, \quad
m_{b}=4.66, \quad m_s=0.095,
\eeq
the mean lifetimes $\tau_{B^0}=1.519\times 10^{-12}~s$ and
$\tau_{B_{s}}=1.512\times 10^{-12}~s$~\cite{Beringer:1900zz,Agashe:2014kda}, and
the values of the Wolfenstein parameters in Refs.~\cite{Beringer:1900zz,Agashe:2014kda}.
%m_{b}=(4.66\pm 0.03), \quad m_s=(0.095\pm0.005),
Choosing the Gegenbauer coefficient as $a_2^{I=0}=0.2$, we extract the parameters
from the LHCb data~\cite{Aaij:2014emv},
\beq
c_1&=&0.900,\quad c_2=0.106, \quad c_3=0.066, \quad c=3.500,\non
\theta_1&=&-\frac\pi2, \quad~~ \theta_2=\frac\pi4, \quad\quad~ \theta_3=0,
\label{para}
\eeq
which correspond to the branching ratios
\beq
{\cal B}(B_s\to J/\psi f_0(980)[f_0(980)\to\pi^+\pi^-])&=&\left(1.15
^{+0.49}_{-0.38}(\omega_{B_s})^{+0.18}_{-0.15}(a^{I=0}_2)^{+0.02}_{-0.01}(m_c)\right)\times10^{-4},\non
{\cal B}(B_s\to J/\psi f_0(1500)[f_0(1500)\to\pi^+\pi^-])&=&\left(1.62
^{+0.54}_{-0.42}(\omega_{B_s})^{+0.28}_{-0.20}(a^{I=0}_2)^{+0.03}_{-0.02}(m_c)\right)\times10^{-5},\non
{\cal B}(B_s\to J/\psi f_0(1790)[f_0(1790)\to\pi^+\pi^-])&=&\left(3.26
^{+0.82}_{-0.71}(\omega_{B_s})^{+0.54}_{-0.38}(a^{I=0}_2)^{+0.05}_{-0.04}(m_c)\right)\times10^{-6},\non
{\cal B}(B^0\to J/\psi f_0(500)[f_0(500)\to\pi^+\pi^-])&=&\left(6.91
^{+3.37}_{-2.16}(\omega_{B})^{+1.58}_{-1.15}(a^{I=0}_2)^{+0.09}_{-0.08}(m_c)\right)\times10^{-6},
\label{pqcd-prediction}
\eeq
with the three errors coming from the variations of $\omega_{B_s}=(0.50\pm0.05)$ GeV~\cite{Ali:2007ff} or
$\omega_{B}=(0.40\pm0.04)$ GeV, $a_2^{I=0}=0.2\pm0.2$, and $m_c=(1.275\pm0.025)$ GeV. It is seen
that the above results are not sensitive to $a_2^{I=0}$, so the central value of
$a_2^{I=0}$ is actually mainly determined by the $B_s\to\pi^+\pi^-\mu^+\mu^-$ data. The errors induced by the
variations of the Wolfenstein parameters and the mean lifetime of the $B^0$ or $B_s$ meson are
tiny, and have been omitted.

We may set the $B_s\to J/\psi f_0(1500)[f_0(1500)\to\pi^+\pi^-]$ contribution to be
$10.0\%$ of the total $B_s\to J/\psi\pi^+\pi^-$ decay rate, which roughly
agrees with the corresponding values in both Solutions I and II presented
by the LHCb Collaboration~\cite{Aaij:2014emv}. The percentages
of the other channels are then given by $70.9\%$ and $2.0\%$
for $B_s\to J/\psi f_0(980)[f_0(980)\to\pi^+\pi^-]$ and
$B_s\to J/\psi f_0(1790)[f_0(1790)\to\pi^+\pi^-]$, respectively,
and support the Solution I data~\cite{Aaij:2014emv}.
Including all the $S$-wave resonances $f_0(980)$, $f_0(1500)$
and $f_0(1790)$ in the scalar form factor, we have the total branching ratio
\beq
{\cal B}(B_s\to J/\psi (\pi^+\pi^-)_{S})=\left(1.62^{+0.69}_{-0.52}
(\omega_{B_s})^{+0.24}_{-0.19}(a^{I=0}_2)^{+0.03}_{-0.02}(m_c)\right)\times10^{-4},
\eeq
to which the interference between $f_0(980)$ and $f_0(1500)$ ($f_0(980)$ and $f_0(1790)$,
$f_0(1500)$ and $f_0(1790)$) contributes
$1.61\times10^{-5}(1.42\times10^{-5}, -2.13\times10^{-6})$.
The $B^0\to J/\psi f_0(500)[f_0(500)\to\pi^+\pi^-]$ branching ratio predicted
in Eq.~(\ref{pqcd-prediction}) is consistent with the data
$(8.8\pm0.5^{+1.1}_{-1.5})\times10^{-6}$ in Ref.~\cite{Aaij:2014siy} and
$(6.4\pm0.8^{+2.4}_{-0.8})\times10^{-6}$ in Ref.~\cite{Aaij:2013zpt}.

For comparisons with other data, we read the branching
ratio ${\cal B}(\bar B_s\to J/\psi\pi^+\pi^-)$ out of
the ratio~\cite{LHCb:2012ae}
\beq
\frac{{\cal B}(\bar B_s\to J/\psi\pi^+\pi^-)}{{\cal B}(\bar B_s\to J/\psi\phi)}=(19.79\pm0.47\pm0.52)\%,
\label{data-lhcb-Bs}
\eeq
with the branching ratio
${\cal B}(\bar B_s\to J/\psi\phi)=(1.07\pm 0.09)\times 10^{-3}$~\cite{Beringer:1900zz,Agashe:2014kda}.
Then a value ${\cal B}(B_s\to J/\psi f_0(980)[f_0(980)\to\pi^+\pi^-])
= (1.48\pm0.14)\times10^{-4}$ is inferred from its percentage in the
total decay rate of $B_s\to J/\psi \pi^+\pi^-$~\cite{LHCb:2012ae}.
This outcome and the Belle data ${\cal B}(B_s\to J/\psi f_0(980)[f_0(980)\to\pi^+\pi^-])=
(1.16^{+0.31}_{-0.19}(stat)^{+0.15}_{-0.17}(syst)^{+0.26}_{-0.18}(N_{B^{(*)}_s\bar B^{(*)}_s}))
\times10^{-4}$~\cite{Li:2011pg} are both compatible with the PQCD prediction in
Eq.~(\ref{pqcd-prediction}).

A ratio of the decay rates,
\beq
R_{f_0/\phi}\equiv\frac{\Gamma(B^0_s\to J/\psi f_0,f_0\to\pi^+\pi^-)}
     {\Gamma(B^0_s\to J/\psi \phi,\phi\to K^+K^-)}=0.252^{+0.046+0.027}_{-0.032-0.033},
\label{lhcb-ratio1}
\eeq
with $f_0$ standing for the $f_0(980)$ meson, was measured by the LHCb
Collaboration~\cite{Aaij:2011fx}, and subsequently confirmed by the CDF
and D0 Collaborations~\cite{Aaltonen:2011nk,Abazov:2011hv}.
We predict this ratio
\beq
R_{f_0/\phi}=0.220^{+0.094+0.034+0.004+0.020}
     _{-0.073-0.029-0.002-0.017},
\eeq
where the first (second, third) error comes from the variation of the shape parameter
$\omega_{B_s}$ (the Gegenbauer moment $a^{I=0}_2$, the charm-quark mass),
and the fourth one is attributed to the uncertainties of
the $B^0_s\to J/\psi\phi$ and $\phi(1020)\to K^+K^-$ decay rates~\cite{Beringer:1900zz,Agashe:2014kda}.
Obviously, our prediction for $R_{f_0/\phi}$ matches the data in
Refs.~\cite{Aaij:2011fx,Aaltonen:2011nk,Abazov:2011hv}.

%%%%%%%%%%%%%%%%%%%%%
%%-----------------------------------------------
\begin{figure}[thb]
\begin{center}
\vspace{-0.5cm}
\centerline{\epsfxsize=9.0cm \epsffile{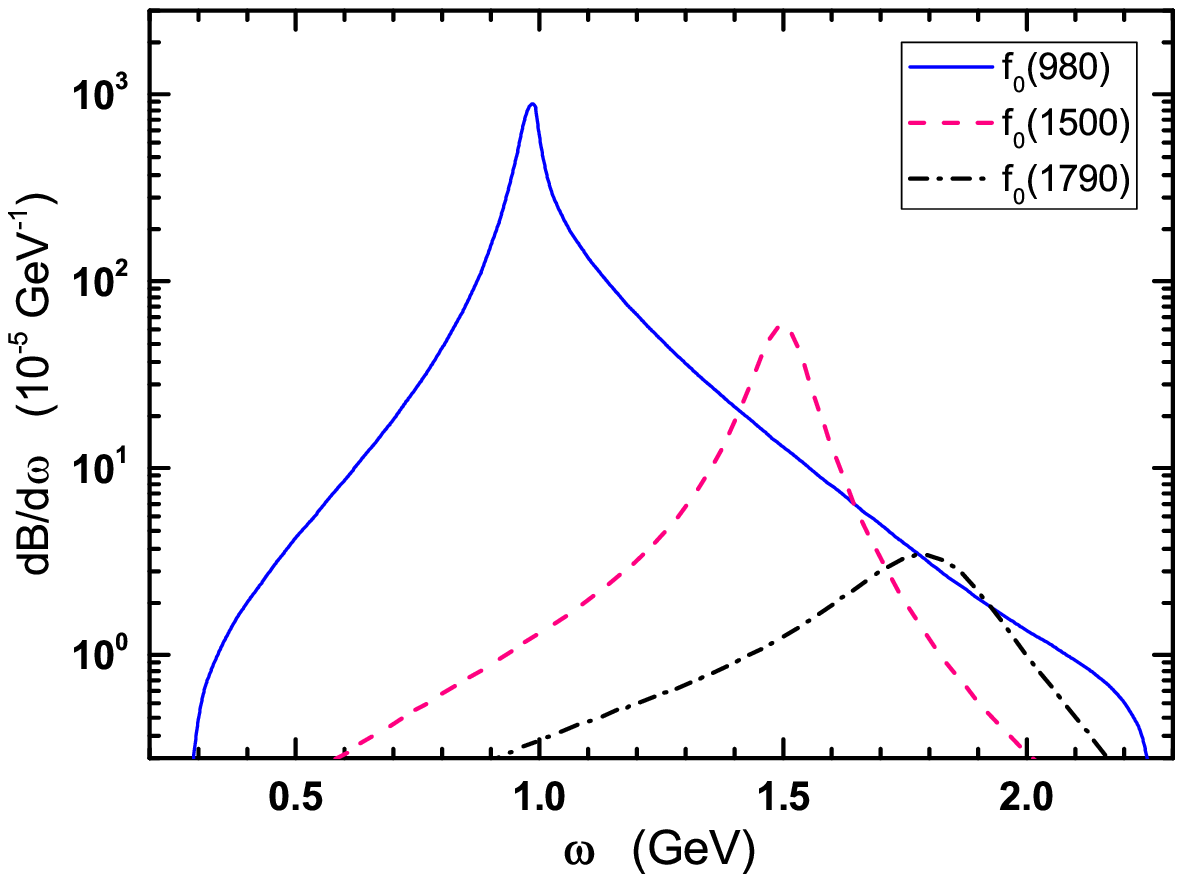}
\epsfxsize=9.0cm \epsffile{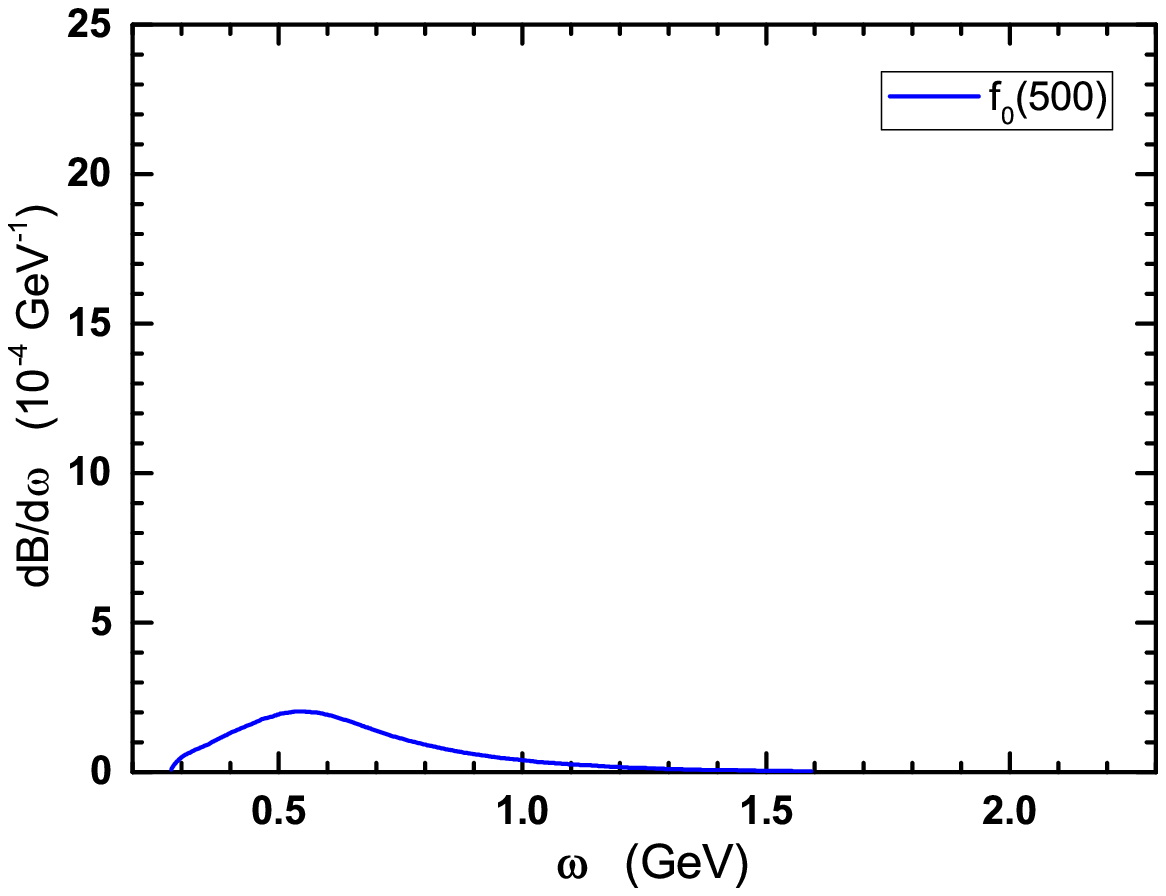}  }
(a)\hspace{9.0cm}(b)
\caption{ Dependencies of the differential branching ratios $d{\cal B}/d\omega$ on the
pion-pair invariant mass for (a) the resonances $f_0(980)$, $f_0(1500)$, and $f_0(1790)$
in the $B_s\to J/\psi\pi^+\pi^-$ decay, and (b) $f_0(500)$ in the
$B^0\to J/\psi\pi^+\pi^-$ decay.}
%\vspace{-0.5cm}
\label{figs-dep}
\end{center}
\end{figure}
%%----------------------------------------------
%%%%%%%%%%%%%%%%%%%%%%%%%

The predicted dependencies of the differential branching ratios $d{\cal B}/d\omega$
on the pion-pair invariant mass
$\omega$ are presented in Fig.~\ref{figs-dep}(a) for the resonances $f_0(980)$, $f_0(1500)$,
and $f_0(1790)$ in the $B_s\to J/\psi\pi^+\pi^-$ decay, and in Fig.~\ref{figs-dep}(b)
for $f_0(500)$ in the $B^0\to J/\psi\pi^+\pi^-$ decay, which can be compared with the data
in Refs.~\cite{Aaij:2014siy,Aaij:2014emv,Aaij:2013zpt,LHCb:2012ae}.
A value of the differential branching ratio at each invariant mass of the pion pair is proportional
to the event number observed by the LHCb Collaboration. We have adjusted the range
of the $y$ axis, such that it is easy to see the coincidence of the the curves
in Fig.~\ref{figs-dep} and the corresponding ones in Refs.~\cite{Aaij:2014siy,Aaij:2014emv}.

%%%%%%%%%%%%%%%%%%%%%%%%%%%%%%%%%%%%%%%%%%%%%%%%%%%%
\begin{figure}\begin{center}
\includegraphics[scale=0.6]{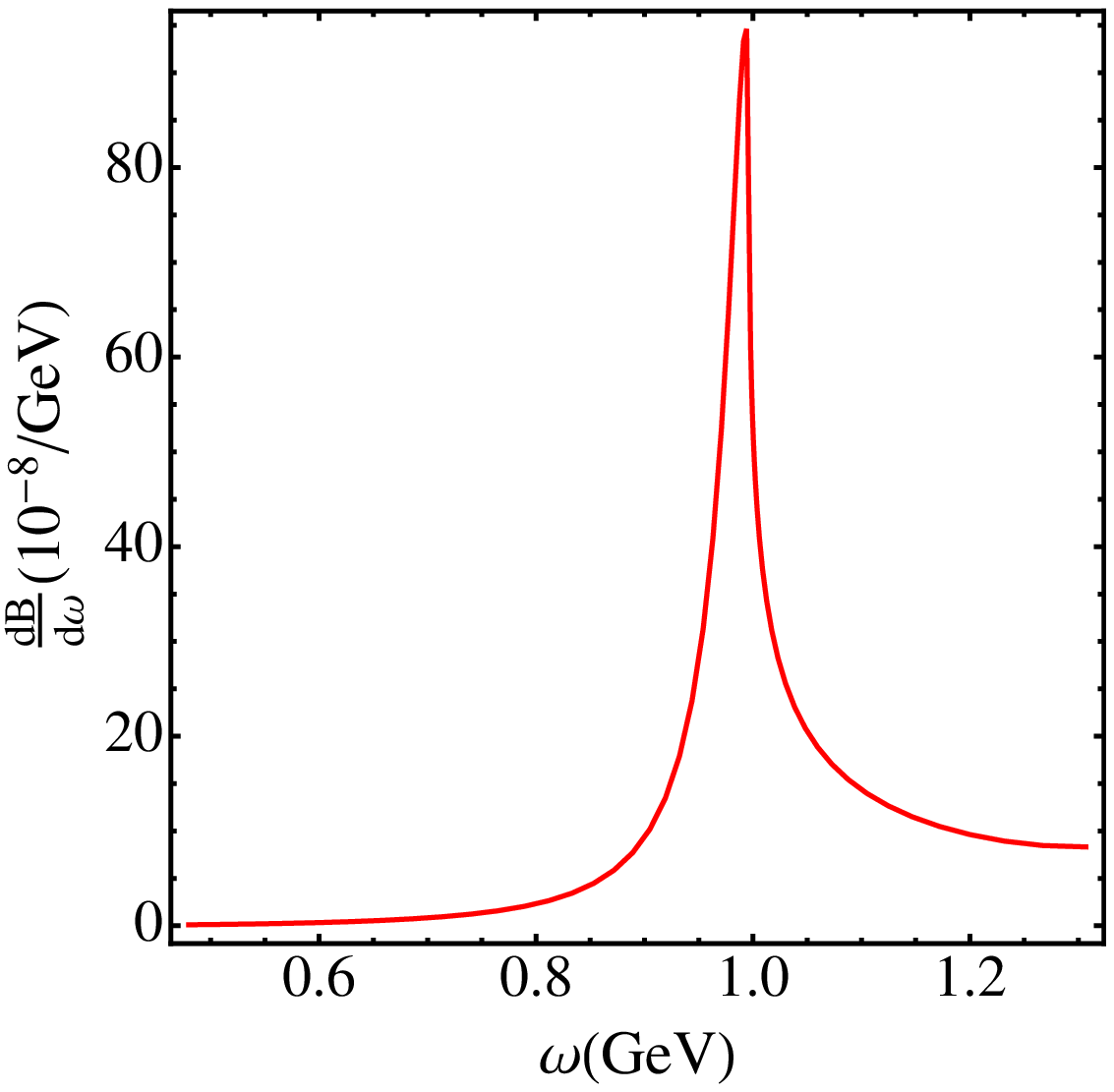}\hspace{1.0 cm}
\includegraphics[scale=0.6]{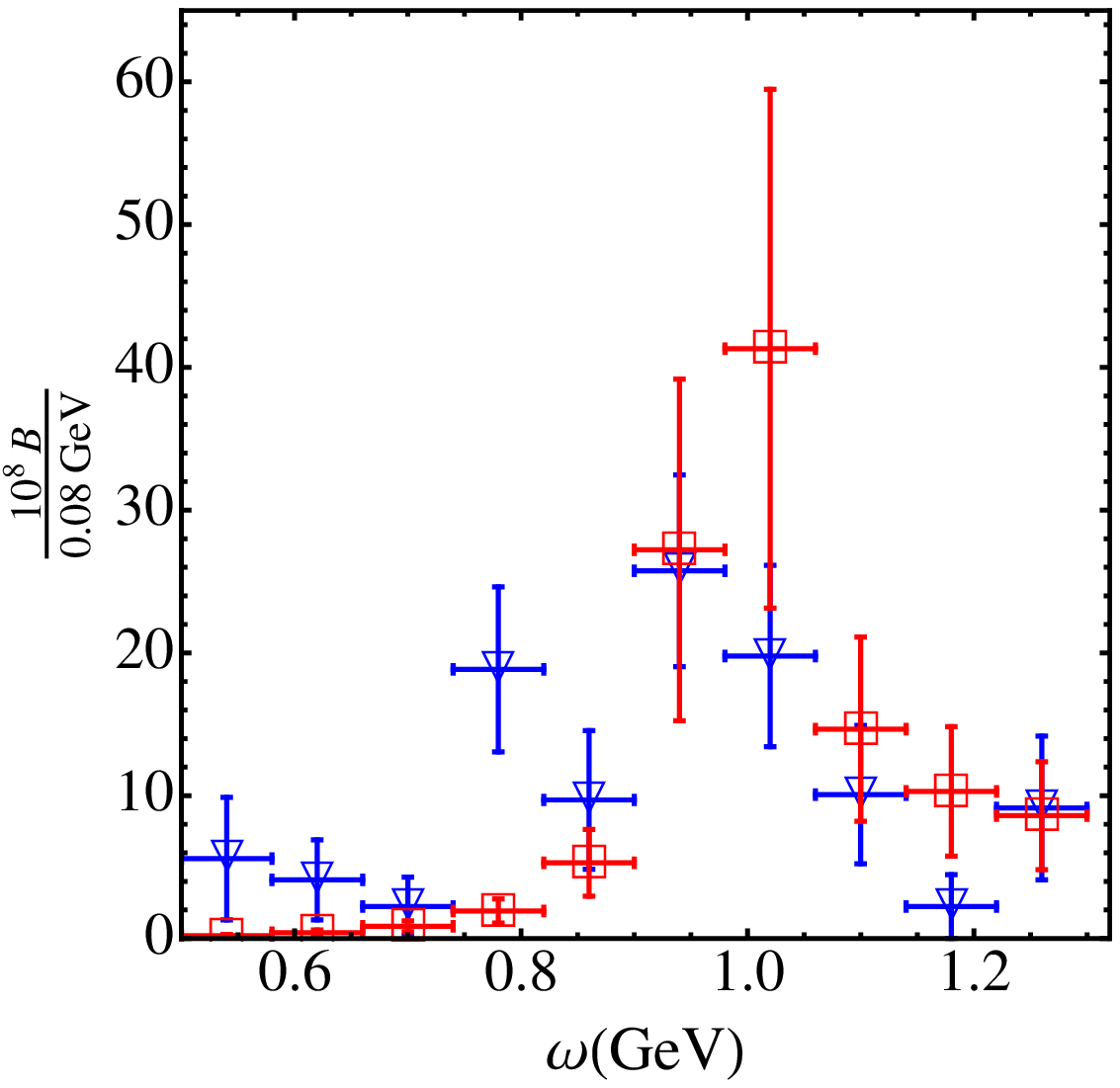}\\
(a)\hspace{8.0 cm}(b)
\caption{(a) Predicted differential branching ratio
for the $B_s\to \pi^+\pi^-\mu^+\mu^-$ decay.
(b) Comparison between the experimental data (with triangle markers)
the theoretical predictions (with square markers) for the differential
branching ratio of the $B_s\to \pi^+\pi^-\mu^+\mu^-$ decay. }
\label{fig:dgammadM2h_the}
\end{center}
\end{figure}
%%%%%%%%%%%%%%%%%%%%%%%%%%%%%%%%%%%%%%%%%%%%%%%%%%%%%

The integrated  branching fraction for the $B_s\to  \pi^+\pi^- \mu^+\mu^-$ decay was
measured by the LHCb as~\cite{Aaij:2014lba}
\begin{eqnarray}
{\cal B}(B_s\to  \pi^+\pi^- \mu^+\mu^-) = (8.6\pm 1.5\pm 0.7\pm 0.7)\times 10^{-8}, \label{eq:Bspipidata}
\end{eqnarray}
where the first two errors are statistical and systematic, respectively, and the
third error is due to the uncertainty from the normalization, i.e. the
$B^0\to J/\psi(\to \mu^+\mu^-) K^*(\to K^+\pi^-)$ branching fraction.
Adopting the parameters of the $S$-wave two-pion distribution amplitudes fixed
in Eq.~(\ref{para}), we calculate Figs.~\ref{fig-fig1}(a) and \ref{fig-fig1}(b), and the
differential branching fraction for the $B_s\to \pi^+\pi^-\mu^+\mu^-$ decay. Our
result is displayed in Fig.~\ref{fig:dgammadM2h_the}(a), which clearly exhibits the
peak arising from the $f_0(980)$ resonance. Integrating over $\omega$, we derive
\begin{eqnarray}
{\cal B}(B_s\to   \pi^+\pi^-  \mu^+\mu^-) = (8.0 \pm 2.4(\omega_{B_s})\pm 2.5(a^{I=0}_2))\times 10^{-8}, \label{eq:theory_Bs_f0_mumu}
\end{eqnarray}
consistent with the data in Eq.\eqref{eq:Bspipidata}, where the theoretical errors
are attributed to the variations of $\omega_{B_s}=(0.50\pm0.05)$ GeV and
$a^{I=0}_2=0.2\pm0.2$. In order to be compared with
the LHCb data~\cite{Aaij:2014lba} carefully, we also present the binned results
in Fig.~\ref{fig:dgammadM2h_the}(b) from $0.5$ GeV to $1.3$ GeV, in which the data
(with triangle markers) have been normalized to the central value
${\cal B}(B_s^0\to \pi^+\pi^-\mu^+\mu^-)=8.6\times 10^{-8}$.
The comparison shows a general agreement  between  our theoretical predictions and the data.
The different around $\omega\sim 0.8$ GeV, if becoming more significant
in the future, may imply the inclusion of the missing $f_0(600)$ resonance into
the parameterized form factor.
It is noticed that the $B_s\to   \pi^+\pi^-  \mu^+\mu^-$ mode is more sensitive to the
parameter $a^{I=0}_2$ than $B_s\to  J/\psi \pi^+\pi^-$, so the comparison in
Fig.~\ref{fig:dgammadM2h_the}(b) is more
crucial for the determination of $a^{I=0}_2$ as stated before.

With the assumption on the dominance of the $f_0(980)\to \pi^+\pi^-$ transition,
the $B_s\to f_0(980) \mu^+\mu^-$ branching fraction was found to be~\cite{Aaij:2014lba}
\begin{eqnarray}
{\cal B}(B_s\to f_0(980)(\to \pi^+\pi^-) \mu^+\mu^-) = (8.3\pm 1.7)\times 10^{-8}. \label{eq:Bs_f0_mumu_data}
\end{eqnarray}
On the other hand, the branching fraction ${\cal B}(B_s\to J/\psi f_0)  =
(1.39\pm0.14)\times 10^{-4}$~\cite{Agashe:2014kda} would indicate
\begin{eqnarray}
{\cal B}(B_s\to f_0(980) \mu^+\mu^-) \sim  6.1 \times 10^{-8},  \label{eq:Bsf0_expec}
\end{eqnarray}
according to the ratio~\cite{Agashe:2014kda}
\begin{eqnarray}
\frac{
{\cal B}( B^-\to K^-\mu^+\mu^-)}{{\cal B}( B^-\to J/\psi K^-) }  =\frac{ (4.49\pm0.23)\times 10^{-7}}{ (1.027\pm 0.031)\times 10^{-3}} \sim 4.4\times 10^{-4},
\end{eqnarray}
if the above ratio was not sensitive to light final-state mesons as in most cases.
The result in Eq.~(\ref{eq:Bsf0_expec}) is consistent with that in
Eq.~(\ref{eq:Bs_f0_mumu_data}), but the central value is lower by about $30\%$.
Both Eqs.~~(\ref{eq:Bs_f0_mumu_data}) and (\ref{eq:Bsf0_expec}) agree with our prediction in Eq.~(\ref{eq:theory_Bs_f0_mumu}) basically.
More precise data from the LHCb and the future KEKB will test our theoretical formalism.

\section{CONCLUSION}

In this work we have studied the contributions from the $S$-wave resonant states
$f_0(980), f_0(1500)$ and $f_0(1790)$ to the $B_s\to J/\psi\pi^+\pi^-$ decay, and from
$f_0(500)$ to the $B^0\to J/\psi\pi^+\pi^-$ decay in the PQCD approach with the
introduction of the $S$-wave two-pion distribution amplitudes. The BW formula for $f_0(500)$,
$f_0(1500)$ and $f_0(1790)$ and the Flatt\'e model for $f_0(980)$ were adopted to describe
the involved timelike scalar form factors, which contain both resonant and nonresonant
dynamics in the two-pion system. The resultant branching fractions, ratio
$R_{f_0/\phi}$, and differential branching fractions in the pion-pair invariant mass
for the relevant resonance channels all agree with the data well. The same two-pion
distribution amplitude associated with the $s\bar s$ component also leads to the differential
branching fraction of the $B_s\to\pi^+\pi^-\mu^+\mu^-$ decay, which is compatible
with the data. It is stressed that the data of both $B_s\to J/\psi\pi^+\pi^-$ and
$B_s\to\pi^+\pi^-\mu^+\mu^-$ modes are required for determining
the timelike scalar form factor and the Gegenbauer coefficient in the $S$-wave
two-pion distribution amplitudes completely.
We conclude that our formalism with the two-hadron distribution amplitudes is successful,
and applicable to other three-body hadronic $B$ meson decays. The extracted $S$-wave
two-pion distribution amplitudes can be employed to predict direct $CP$ asymmetries of
the decays with intermediate $S$-wave resonances in various localized regions of the two-pion
phase space.

%-----------------------vv----------------------------------------%
\begin{acknowledgments}
We thank M. Nakao, B. Pire and E. Oset  for useful discussions. This work was supported in part
by the Ministry of Science and Technology of R.O.C. under Grant No.
NSC-101-2112-M-001-006-MY3, by National Science Foundation
of China under Grant Nos. 11235005, 11375208, and 1122851, by a key laboratory grant from
the Office of Science and Technology, Shanghai
Municipal Government (No. 11DZ2260700),  by Shanghai Natural  Science Foundation
under Grant No. 15ZR1423100, and by China Postdoctoral Science Foundation.
\end{acknowledgments}

%---------------------------------------------------------------%
%---------------------------------------------------------------%
\appendix

\section{Decay amplitudes}

In this Appendix we present the PQCD factorization formulas for the diagrams in
Fig.~\ref{fig-fig1}. The $J/\psi$ vector meson distribution amplitudes with longitudinal
polarization read as
\begin{eqnarray}
\Phi^L_{J/\psi}=\frac{1}{\sqrt{2N_c}}\left[m_{J/\psi}\epsl_L\psi^L+\epsl_L\psl_3\psi^t\right],
\end{eqnarray}
with the longitudinal polarization vector
\beq
\epsilon_L=\frac{m_B}{\sqrt2m_{J/\psi}}\left(-r^2,(1-\eta),0_\rmt\right).
\eeq
The twist-2 distribution amplitude $\psi^L$ and the twist-3 distribution amplitude
$\psi^t$ take the forms \cite{Bondar:2004sv}
\begin{eqnarray}
\psi^L(x)&=&9.58\frac{f_{J/\psi}}{2\sqrt{2N_c}}x(1-x)\left[\frac{x(1-x)}{1-2.8x(1-x)}\right]^{0.7},\non
\psi^t(x)&=&10.94\frac{f_{J/\psi}}{2\sqrt{2N_c}}(1-2x)^2\left[\frac{x(1-x)}{1-2.8x(1-x)}\right]^{0.7},
\end{eqnarray}
with the $J/\psi$ meson decay constant $f_{J/\psi}$.
The distribution amplitudes and the relevant parameters for the $B^0$ and $B_s$ mesons
can be found in Ref.~\cite{Xiao:2011tx}.

The analytic formula for the $B^0_{(s)}\to J/\psi\pi^+\pi^-$
decay amplitude is given by
\beq
\mathcal{A}({B^0_{(s)}\to J/\psi\pi^+\pi^-})=V^*_{cb}V_{cd(cs)}\left(F^{LL}+M^{LL}\right)
-V^*_{tb}V_{td(ts)}\left(F^{\prime LL}+F^{LR} +M^{\prime LL} +M^{SP} \right).
\eeq
For simplicity, we denote the distribution
amplitudes $\Phi_{v\nu=-}^{I=0}(z,\zeta,w^2)$
$[\Phi_{s}^{I=0}(z,\zeta,w^2), \Phi_{t\nu=+}^{I=0}(z,\zeta,w^2)]$
by $\phi_0$ $(\phi_s, \phi_\sigma)$ below.
The amplitudes for the $B$ ($B^0$ or $B_s$) meson transition into two pions from Fig.~\ref{fig-fig1}(a)
and Fig.~\ref{fig-fig1}(b) are written as
\beq
 F^{LL}&=&8\pi C_F m^4_B f_{J/\psi}\int_0^1 dx_B dz
\int_0^\infty b_B db_B b db \phi_B(x_B,b_B)\non
&&\times\bigg\{\sqrt{\eta(1-r^2)}\left[((1-2z)(1-\eta)-r^2(1-2z(1-\eta)))(\phi_s+\phi_\sigma)
+2r^2\phi_\sigma\right]\non
&&+\left[(1+z)(1-\eta)-r^2(1+2z(1-\eta))\right]\phi_0
 \bigg]a_1(t_a)E_e(t_a)h_a(x_B,z,b_B,b)\non
&&+\bigg[2\sqrt{\eta(1-r^2)}\left[1-\eta-r^2(1-x_B)\right]\phi_s\non
&&-(1-r^2)\left[(1-\eta)\eta + r^2(x_B-\eta)\right]\phi_0\bigg]a_1(t_b)E_e(t_b)h_b(x_B,z,b_B,b)
\bigg\},\\
F^{\prime LL}&=&F^{LL}|_{a_1\to a_2},\\
F^{LR}&=&F^{LL}|_{a_1\to a_3},
\label{exp-F-LL}
\eeq
with the Wilson coefficients $a_1=C_1+C_2/N_c$,
$a_2=C_3+C_4/N_c+C_9+C_{10}/N_c$, and $a_3=C_5+C_6/N_c+C_7+C_8/N_c$.
The spectator diagrams
in Fig.~\ref{fig-fig1}(c) and Fig.~\ref{fig-fig1}(d) yield the amplitudes
\beq
M^{LL}&=&-32\pi C_F m^4_B/\sqrt{6} \int_0^1 dx_B dz dx_3
\int_0^\infty b_B db_B b_3 db_3 \phi_B(x_B,b_B)\non
&&\times\bigg\{
\bigg[\sqrt{\eta(1-r^2)}[2(1-\eta)(z(1-r^2)+r^2(1-x_3))\phi_\sigma\non
&&-(z(1-\eta)(1-r^2)+r^2x_B)(\phi_s+\phi_\sigma)]\psi^L
+(1-r^2-\eta) \non
&&\times\big[\left((1-x_3-x_B)(1-r^2)+\eta(x_3(1-2r^2)-(1-r^2)(1-z)+r^2)\right)\psi^L \non
&&+rr_c(1-\eta)\psi^t \big]\phi_0\bigg]C_2(t_c)E_n(t_c)h_c(x_B,z,x_3,b_B,b_3)\non
&&-\bigg[\sqrt{\eta(1-r^2)}\big[(2r^2(x_B-x_3(1-\eta))\psi^L+4rr_c(1-\eta)\psi^t)\phi_\sigma\non
&&-(r^2x_B+z(1-\eta)(1-r^2))(\phi_s+\phi_\sigma)\psi^L\big]\non
&&-(1-r^2-\eta)\left[(x_B-z-x_3(1-\eta)+r^2(z-x_3))\psi^L+rr_c(1-\eta)\psi^t\right]\phi_0\bigg]\non
&&\times C_2(t_d) E_n(t_d)h_d(x_B,z,x_3,b_B,b_3) \bigg\},\\
M^{\prime LL}&=&M^{LL}|_{C_{2}\to a_{4}},
\label{exp-M-LL}
\eeq
\beq
M^{SP}&=&32\pi C_F m^4_B/\sqrt{6} \int_0^1 dx_B dz dx_3
\int_0^\infty b_B db_B b_3 db_3 \phi_B(x_B,b_B)\non
&&\times\bigg\{
\bigg[-\sqrt{\eta(1-r^2)}\big[2r^2((1-x_3)(1-\eta)-x_B)\psi^L-4rr_c(1-\eta)\psi^t\big]\phi_\sigma\non
&&+\sqrt{\eta(1-r^2)}\left[z(r^2-1)(1-\eta)-r^2x_B\right](\phi_s+\phi_\sigma)\psi^L\non
&&+(1-r^2-\eta)\big[((1-x_3)(1+r^2-\eta)+z(1-r^2)-x_B)\psi^L\non
&&-rr_c(1-\eta)\psi^t\big]\phi_0\bigg]a_5(t_c)E_n(t_c)h_c(x_B,z,x_3,b_B,b_3)\non
&&+\bigg[\sqrt{\eta(1-r^2)}\left[2(\eta-1)(r^2(x_3-z)+z)\phi_\sigma+(z(1-r^2)(1-\eta)+r^2x_B)(\phi_s+\phi_\sigma)\right]\psi^L
\non
&&+(1-r^2-\eta)\big[\left((1-r^2)(x_B-z\eta)+x_3(r^2(1-2\eta)-(1-\eta))\right)\psi^L \non
&&-rr_c(1-\eta)\psi^t \big]\phi_0 \bigg]a_5(t_d)E_n(t_d)h_d(x_B,z,x_3,b_B,b_3)
 \bigg\},
\label{exp-M-SP}
\eeq
with the mass ratio $r_c=m_c/m_B$, the notations
$\psi^{L,t}\equiv\psi^{L,t}(x_3)$, and the Wilson coefficients $a_4=C_4+C_{10}$ and $a_5=C_6+C_8$.

The hard functions
$h_a, h_b, h_c$ and $h_d$ are written as
\beq
h_a(x_B,z,b_B,b)&=&
\big[\theta(b-b_B)I_0(m_B\sqrt{z(1-r^2)}b_B)K_0(m_B\sqrt{z(1-r^2)}b)
+(b\leftrightarrow b_B) \big]\non
&&\times K_0(m_B\sqrt{\alpha}b_B)S_t(z),
\non
h_b(x_B,z,b_B,b)&=&\left\{ \begin{array}{ll}
 \left[\theta(b-b_B)I_0(m_B\sqrt{\kappa}b_B)K_0(m_B\sqrt{\kappa}b)
 +(b\leftrightarrow b_B) \right],\quad\quad\quad\quad \kappa\geq 0 \\
 \frac{i\pi}{2}\left[\theta(b-b_B)J_0(m_B\sqrt{|\kappa|}b_B)H_0^{(1)}(m_B\sqrt{|\kappa|}b)
  +(b\leftrightarrow b_B) \right], ~ \kappa< 0 \\
\end{array} \right.\non
&& \times K_0(m_B\sqrt{\alpha}b)S_t(|x_B-\eta|) ,\non
h_c(x_B,z,x_3,b_B,b_3)&=&
 \left[\theta(b_B-b_3)I_0(m_B\sqrt{\alpha}b_3)K_0(m_B\sqrt{\alpha}b_B)
 +(b_B\leftrightarrow b_3) \right]\non
&&\times \left\{ \begin{array}{ll}
K_0(m_B\sqrt{\beta}b_3),\quad\quad\quad\quad \beta\geq 0   \\
\frac{i\pi}{2}H_0^{(1)}(m_B\sqrt{|\beta|}b_3),\quad\quad \beta< 0  \\
\end{array} \right.
\non
h_d(x_B,z,x_3,b_B,b_3)&=&h_c(x_B,z,1-x_3,b_B,b_3),\label{hard}
\eeq
with the factors $\alpha=(1-r^2)x_B z$, $\kappa=(1-r^2)(x_B-\eta)$, and
$\beta=r_c^2-(z(1-r^2)+r^2(1-x_3))((1-\eta)(1-x_3)-x_B)$, and the Hankel function
$H_0^{(1)}(x)=J_0(x)+iY_0(x)$. In the above expressions the threshold resummation factor $S_t(x)$ follows the
parametrization in Ref.~\cite{Kurimoto:2001zj}
\beq\label{eq-def-stx}
S_t(x)=\frac{2^{1+2c}\Gamma(3/2+c)}{\sqrt{\pi}\Gamma(1+c)}[x(1-x)]^c,
\eeq
with the fitted parameter~\cite{Li:2009pr}
\begin{eqnarray}
c=0.04 Q^2-0.51Q+1.87,
\end{eqnarray}
and $Q^2=m^2_{B}(1-r^2)$ or $Q^2=m^2_{B}(1-q^2/m_{B}^2)$.
%as the energy carried by the two-hadron system in the light-cone coordinate.

The evolution factors $E_e(t)$ and $E_n(t)$ in the factorization formulas are given by
\beq
E_e(t)&=&\alpha_s(t) \exp[-S_B(t)-S_M(t)],\non
E_n(t)&=&\alpha_s(t) \exp[-S_B(t)-S_M(t)-S_{J/\psi}(t)]|_{b=b_B},
\eeq
in which the Sudakov exponents are defined as
\beq
S_B&=&s\left(x_B\frac{m_B}{\sqrt2},b_B\right)+\frac53\int^t_{1/b_B}\frac{d\bar\mu}{\bar\mu}
\gamma_q(\alpha_s(\bar\mu)),\\
S_M&=&s\left(z(1-r^2)\frac{m_B}{\sqrt2},b\right)+s\left((1-z)(1-r^2)\frac{m_B}{\sqrt2},b\right)+
2\int^t_{1/b}\frac{d\bar\mu}{\bar\mu}
\gamma_q(\alpha_s(\bar\mu)),\\
S_{J/\psi}&=&s\left(x_3(1-\eta)\frac{m_B}{\sqrt2},b_3\right)
+s\left((1-x_3)(1-\eta)\frac{m_B}{\sqrt2},b_3\right)+
2\int^t_{1/b_3}\frac{d\bar\mu}{\bar\mu}
\gamma_q(\alpha_s(\bar\mu)),
\eeq
with the quark anomalous dimension $\gamma_q=-\alpha_s/\pi$. The explicit expression
of the function $s$ can be found, for example, in Appendix A of Ref.~\cite{Wang:2012ab}.
The hard scales $t_a, t_b, t_c$ and $t_d$ involved in Figs.~\ref{fig-fig1}(a)-\ref{fig-fig1}(d)
are chosen as
\beq
t_a&=&\max\left\{m_B\sqrt{z(1-r^2)}, 1/b, 1/b_B\right\},\quad\quad
t_b =\max\left\{m_B\sqrt{|\kappa|}, 1/b, 1/b_B\right\},\non
t_c&=&\max\left\{m_B\sqrt{\alpha},m_B\sqrt{|\beta|},1/b_B, 1/b_3\right\},~~~
t_d =t_c|_{x_3\to(1-x_3)}.
\eeq

The PQCD factorization formulas for the $B_s\to\pi^+\pi^-$ form factors with
the $S$-wave two-pion distribution
amplitudes are written as
\begin{eqnarray}
  {\cal F}_{0}(q^2)&=&8 \pi C_F m_{B_s}^2 \int_0^1
        dx_{B}dz\int_0^{\infty}b_Bdb_Bbdb \phi_{B_s}(x_{B},b_B)\nonumber\\
        &&\times\left\{\big[\eta(z\eta+1)\phi_{0}+\sqrt{\eta(1-r^2)}\eta(1-2z)\phi_{\sigma}
        +\sqrt{\eta(1-r^2)}(-2z\eta-\eta+2)\phi_{s}\big]\right.\nonumber\\
        &&\left.\times h_e(x_{B},z\eta,b_B,b)\alpha_s(t_e^1)\mbox{exp}[-S_{B}(t_e^1)-S_{M}(t_e^1)]S_t(z)\right.\nonumber\\
        &&\left.+2\sqrt{\eta(1-r^2)}\eta \phi_{s}h_e(z,x_{B}\eta,b,b_B)\alpha_s(t_e^2)\mbox{exp}[-S_{B}(t_e^2)-S_{M}(t_e^2)]S_t(x_{B})\right\},\label{eq:f0}\\
 %------------------------------------------------------------------------------------------------------------------
  {\cal F}_{1}(q^2) &=&8 \pi C_F m_{B_s}^2 \int_0^1
        dx_{B}dz\int_0^{\infty}b_Bdb_Bbdb \phi_{B_s}(x_{B},b_B)\nonumber\\
        &&\times\left\{\big[(z\eta+1)\phi_{0}-\sqrt{\eta(1-r^2)}(1+2z-2/\eta)\phi_{\sigma}
        -\sqrt{\eta(1-r^2)}(2z-1)\phi_{s}\big]\right.\nonumber\\
        &&\left.\times h_e(x_{B},z\eta,b_B,b)\alpha_s(t_e^1)\mbox{exp}[-S_{B}(t_e^1)-S_{M}(t_e^1)]S_t(z)\right.\nonumber\\
        &&\left.+2\sqrt{\eta(1-r^2)}\phi_{s}h_e(z,x_{B}\eta,b,b_B)\alpha_s(t_e^2)\mbox{exp}[-S_{B}(t_e^2)-S_{M}(t_e^2)]S_t(x_{B})\right\},\\
 %---------------------------------------------------------------------------------------------------------------------
 {\cal F}_{T}(q^2) &=&8 \pi C_F m_{B_s}^2 (1+\sqrt{\eta(1-r^2)})\int_0^1 dx_{B}dz \int_0^{\infty} b_B db_B b db \phi_{B_s}(x_{B},b_B)  \nonumber\\
     && \times\left\{ \big[\sqrt{\eta(1-r^2)}(-z)\phi_{s}+\phi_{0}+\sqrt{\eta(1-r^2)}(z+2/\eta)\phi_{\sigma}\big]\right.\nonumber\\
        &&\left.\times h_e(x_{B},z\eta,b_B,b)\alpha_s(t_e^1)\mbox{exp}[-S_{B}(t_e^1)-S_{M}(t_e^1)]S_t(z)\right.\nonumber\\
        &&\left.+2\sqrt{\eta(1-r^2)}\phi_{s}
        h_e(z,x_{B}\eta,b,b_B)\alpha_s(t_e^2)\mbox{exp}[-S_{B}(t_e^2)-S_{M}(t_e^2)]S_t(x_{B})\right\}.\label{eq:fT}
 \end{eqnarray}
The above expressions are similar to those for
$B$ meson decays into a scalar meson~\cite{Li:2008tk} except the replacement of masses and
distribution amplitudes. The definitions of the hard function $h_e$ and the hard scales
$t_e^{1,2}$ can also be found in Ref.~\cite{Li:2008tk}.
Note that the factors $S_t$ in the second terms depend only on
$x_B$, instead of $x_B-\eta$ as in Eq.~(\ref{hard}), because we considered the small $\eta$
region for the $B_s\to\pi^+\pi^-\mu^+\mu^-$ decay here.

%========================= reference=========================%

\end{document}